\documentstyle[preprint,aps]{revtex}
\tightenlines
\begin{document}
\preprint{UM-P-98/46, RCHEP-98/12}
\draft
\title{Black holes with magnetic charge and quantized mass}

\author{A.Yu.Ignatiev, G.C.Joshi and Kameshwar C.Wali}
\address{Research Centre for High Energy Physics, School of Physics, 
         University of Melbourne,
Parkville 3052, Victoria, Australia\\
Department of Physics, Syracuse University, Syracuse, NY 13210}
\date{Invited paper for the first editorial volume of the book series
"Contemporary Fundamental Physics" by the Nova Science Publishers}
\maketitle
\begin{abstract}
We examine the issue of magnetic charge quantization in the presence
of black holes. It is pointed out that quantization of magnetic charge
can lead to the mass quantization for magnetically charged black holes.
We also discuss some implications for the experimental searches of
magnetically charged black holes.
\end{abstract}
The purpose of this note is to draw attention to some aspects of the
interplay between gravity and the quantization of magnetic charge.
The issue we are going to address is not only of fundamental
theoretical importance but also can be relevant for the experimental
searches of monopoles.

Let us view the problem of magnetic charge from two perspectives.
First, the magnetic charge $g_{RN}$ of the extreme Reissner-Nordstrom 
black hole is related to its mass according to the equation,
\begin{equation}
\label{RN}
g_{RN}=\sqrt{G}M,
\end{equation}
where $M$ is the mass of the black hole. This relation follows from an 
exact solution of Einstein-Maxwell equations \cite {mp,hh} 
, and is a purely classical result with no appeal 
to quantum mechanics. Similarly, an extension to coupled Einstein-
Yang-Mills-Higgs system leads to non-abelian and abelian black holes 
with magnetic charges \cite{vw}. 
References to earlier and related work are contained in this paper. 
More recently, generalizations to slowly rotating non-Abelian 
black holes have been reported by M.S. Volkov and N. Straumann
\cite{vs}.
Second, the Dirac quantization condition \cite{Dirac} reads:
\begin{equation}
\label{D}
g_D={n\hbar c \over 2e}.
\end{equation}
We use the Gauss system of units where $e^2/\hbar c =\alpha$.

Now, what is the relationship between the Eqs.(\ref{RN}) and
(\ref{D})? Should we equate the left-hand sides of the above formulas?
Or are they totally independent of each other? Surprisingly, this
supposedly simple question, to the best of our knowledge, does not
have a straightforward answer. What we really obtain here is
a dilemma: on the one hand, we can insist that the Dirac quantization
condition derived in flat space-time, should be valid also in curved 
space-time in the presence of a black hole; on the other hand, we 
may conjecture that there may exist some mechanism of evading the 
Dirac quantization condition in the regions with strong gravitational 
fields. Both these alternatives have interesting consequences. Which 
of them is realized in Nature seems to be an open question at present. 
For some particular cases it has been claimed that the Dirac quantization
condition remains intact in the presence of gravity \cite{curdir}.
However, there is 
no general proof that we know of, establishing the validity of the 
Dirac quantization condition in arbitrary curved space-time.
Neither there are any models that would definitely establish a 
possibility of its violation. Therefore, we examine qualitatively 
        each scenario and discuss the consequences.

If we assume that the Dirac quantization condition is valid in the
presence of gravity then we can equate the left-hand sides of
Eqs.(\ref{RN}) and (\ref{D}) to obtain the equation
\begin{equation}
M={n\hbar c \over \sqrt{G}2e}=n{M_{Pl} \over 2\sqrt{\alpha}} .
\end{equation}

This says that the masses of Reissner-Nordstrom black holes are 
         quantized
and the minimum possible mass for the extreme magnetically charged
black holes is given by
\begin{equation}
M_{min}=5.85 M_{Pl},
\end{equation}                                
with an equi-spaced mass spectrum analogous to the energy spectrum
of a quantum simple harmonic oscillator. Whether such black holes are 
the final stages of more massive black holes undergoing Hawking radiation and 
reaching a stable state is an interesting question. Anyway, theory 
        aside,  
the range of magnetic monopole masses around the point $5.85 M_{Pl}$ 
         should 
be considered seriously in the monopole search
experiments.

Now, let us consider the other alternative, namely, the possible 
         violation of
the Dirac quantization condition in the presence of gravity. Not 
         knowing
the precise nature of its violation assume that the magnetic charge of a 
Reissner-Nordstrom black hole obeys only one constraint,
        that of Eq.(\ref{RN}), being
unrestricted by Eq.(\ref{D}). Thus, we have 
magnetic monopoles with "dequantized" magnetic charge determined by the 
mass \cite{BPS}.  Thus there is no restriction on
the range of allowed magnetic charges. However, let us confine our 
        attention
to charges  between zero
and one Dirac unit, $g=68.5e$ and note that  very small magnetic charges
correspond to the black holes with the mass much less than the Planck
mass for which the Compton wavelength ${\hbar \over Mc}$ is longer
than the Schwarzschild radius ${2GM \over c^2}$. Therefore, the
effects of quantum gravity should be quite important for such
black holes. For this reason it would be safer to leave very
small magnetic charges from our consideration. A particularly
interesting case from the point of view of symmetry between electricity 
and magnetism, is $g=e$. This equality of magnetic charge of the black 
        hole and
the electric charge of the electron would occur for the black hole
mass
\begin{equation}
M=\sqrt{\alpha}M_{Pl}=0.08M_{Pl}
\end{equation}
Another special case of interest is the black hole with the Planck mass
for which the magnetic charge is
\begin{equation}
g_{Pl}={e \over \sqrt{\alpha}}=11.7e,
\end{equation}
so $g_{Pl}=1$ in dimensionless units ($\hbar=c=1$). Black holes with
the Planck mass have been called maximons in \cite{m}.

In this context, it is interesting to recall that the reported monopole 
observations of Ehrenhaft \cite{e} corresponded to the values of magnetic 
charge {\em less} than the standard Dirac charge of $68.5e$, and hence 
they were given less credibility on the ground that they violated Dirac 
quantization condition.

Are there any arguments in favour of the violation of the Dirac
quantization condition by the gravitational effects? Originally, the
Dirac quantization condition appeared as the necessary condition for
the unobservability  of the Dirac string. The Dirac string is the line
of a singularity in  the vector potential describing the monopole. It
starts at the monopole location and
stretches to infinity. Analytically it can be written, for example,
as
\begin{mathletters}
\label{F}
\begin{eqnarray}
   &A^D_r =\ A^D_\theta = 0,&  \label{F2} \\   & A^D_\varphi = 
        \frac{g}{r} \tan {\frac{\theta}{2}}.& \label{F3}
\end{eqnarray}
\end{mathletters}
The magnetic field corresponding to that vector potential is also
singular and has the form \cite{w} (here, {\bf n} is the unit 
vector along the string, i.e. opposite the z-axis):
\begin{equation}
\nabla \times {\bf A}^{D} = {g \over r^3}{\bf r} + {\bf n}4\pi g \theta 
         (-z) 
\delta (x) \delta (y)
.
\end{equation}
Note that to obtain the above equation rigorously one has to treat all
quantities involved in the sense of distributions rather than regular
functions \cite{st}.
Therefore, the string makes a singular contribution to the energy
momentum tensor of the electromagnetic field. A natural question then
arises: can such a string be observed by its gravitational effects?
The gravitational field of a gauge string has been studied in 
\cite{v}
(A gauge string has a finite mass $\mu$ per unit length whereas a
Dirac string has {\em infinite} line density). It has been found that
the spacetime around such a string is locally flat, but globally it is
similar to a cone. It means that the range of the polar angle
$\phi$ (measured around the string) is not $2\pi$ but  $0 \leq \phi
\leq 2\pi(1-4G\mu)$. This angular deficit has been called a conical
singularity. The presence of conical singularity leads to several
observable effects: double images of sources located behind the
string, anisotropy of the background radiation, and the creation of
sheets of matter in the wake of the string.
These results strongly suggest that the Dirac string {\em may} become
observable by its gravitational effects {\em regardless of the value
of the magnetic charge}. \footnote{Note that there exists some evidence
that the string may become visible even in flat space. In \cite{lp} 
the motion of a monopole in the magnetic field of an external electric
current
was analyzed. Roughly speaking, the physical results depend on how many
times the monopole string winds up around the current line. Another 
configuration where the strings turn out to be a problem has been
considered in \cite{lk}: a charged particle in the magnetic field
of a Dirac monopole {\em line}. Further problems
related to the Dirac quantization condition and existence of strings have 
been raised recently in the context of quantum field theory with monopoles,
see e.g. \cite{he,we,milton}.It is not clear whether or not the 
stringless formalisms of Wu-Yang and others \cite{wy,t,fs} based on the
fibre  
bundle theory
 would help in solving these problems.} 
Thus we are led to conjecture that in the
presence of gravitation it may become irrelevant whether the
quantization condition for the Dirac monopole is fulfilled or
violated. (Of course, an alternative interpretation of the same result
would be to claim that the whole concept of the Dirac monopole is
inconsistent with gravity.)

Let us now turn to the situation with the t'Hooft-Polyakov type of
monopoles. For these monopoles there exist two very different
pictures: one uses a gauge with the ``hedgehog'' configuration of the
scalar field; this gauge is completely free of any singularity lines
(non-singular gauge). The other approach is very close to the picture
of a Dirac monopole because of the presence of a string similar to the
Dirac string (string gauge). These two pictures are believed to be
physically equivalent because there exist a (singular) gauge
transformation from one to the other. Therefore, if we choose
to work in the string gauge we can again require the string to be
unobservable and thus obtain the quantization condition.

Now, we would like to think of the monopole coupled to gravity. One
of the fundamental questions is: will the two gauges remain physically
equivalent? If they are then we can use again the same line of argument 
as above. Consequently, we can again conjecture that the quantization 
condition may become irrelevant in the presence of gravity. On the other 
         hand, 
the non-singular gauge and the string gauge may
become non-equivalent once the gravity is switched on, in which case we 
        do not 
know the form the quantization condition takes. In other words, 
at the present state of our knowledge, arbitrary magnetic charges 
unconstrained by the Dirac quantization condition seem not to be
ruled out by sound theoretical arguments. This should be kept in mind in 
experimental searches. 

We now turn to the analysis of the experimental constraints on the
flux of magnetically charged black holes (``black poles'' for short).
The most stringent constraints come from the results of non-
accelerator experiments devoted to the search of superheavy magnetic
monopoles arising in Grand Unified Theories of strong, weak, and
electromagnetic (but not including gravitational) interactions. These
searches can be divided into two main classes: the first uses the
effect of current induction caused by the monopole magnetic field and
the second is based upon the ionizing properties of monopoles. The
induction experiments give the most direct upper limits on the
monopole flux because they are independent of the unknown
characteristics of the monopoles such as its mass and velocity. The
best upper bounds on the monopole flux $F$ from induction experiments
are:
\begin{equation}
F < 4.4\times 10^{-12} cm^{-2}sr^{-1}s^{-1} \cite{g91},
\end{equation}
\begin{equation}
F < 7.2\times 10^{-13} cm^{-2}sr^{-1}s^{-1} \cite{h91}
\end{equation}
\begin{equation}
F < 5\times 10^{-14} cm^{-2}sr^{-1}s^{-1} \cite{c86}
\end{equation}
These constraints can be taken over for the case of {\em quantized}
black holes regardless of their mass and velocity. As for the {\em
dequantized} black holes, however, the situation is different. The
reason is that the magnetic flux through a superconducting loop is
quantized and can change only in multiples of magnetic flux quantum
$\phi^0={1 \over 2}(4\pi \times {\hbar c \over 2e})$. Therefore, it is
unclear to what extent the superconductive induction experiments are
sensitive to the dequantized magnetic charges. A sufficient condition
for a magnetic charge $g$ to be detected in the induction experiment
reads
\begin{equation}
\label{s}
4\pi g={\phi_0 \over n_l},
\end{equation}
where $n_l$ is a geometric factor which is specific for the loop
configuration adopted in a particular experiment. For instance, for
the original Cabrera's detector $n_l=4$ since there are 4
superconducting loops connected in a series. An interesting problem is
how to design an experimental setup (or use the existing setups) in
such a way that it would be capable of detecting {\em both} quantized
{\em and dequantized} magnetic charges (including the charges not
satisfying Eq.~(\ref{s})).

Let us now turn to the experiments that use the ionizing properties of
monopoles. We consider first the case of quantized black holes,
assuming that the black holes with $n \geq 1$ are stable.
The ionization experiments are much more sensitive to the theoretical
input than the induction experiments. In particular, their sensitivity
crucially depends on the monopole's velocity: the monopole would be
undetectable if its velocity were below the threshold, somewhere
between $10^{-3}c$ and $10^{-4}c$. To find out the typical monopole
velocity, we need to find the greater of the two characteristic
velocities: the first is the velocity with which the monopole enters
the Galaxy ($\sim 10^{-3}c$) and the second is the typical monopole
velocity due to its acceleration by the galactic magnetic field,
given by
\begin{equation}
v_n \simeq 3\times 10^{-3}c({10^{16}GeV \over M_n})^{1/2}.
\end{equation}
                                 
The quantized magnetically charged black holes with masses $M \agt
6\times 10^{19}$ GeV
can be viewed as {\em ultraheavy} monopoles (as contrasted with {\em
superheavy} GUT monopoles with the mass of the order of $10^{16}$
GeV). It is readily seen that the total spectrum of the quantized
black poles can be divided into two parts: 1)``low-lying'', low charge
black poles, corresponding to $n \alt 600$ and 2)``very heavy '',
heavily charged black poles'' with $n \agt 600$. For the low-lying
black poles the galactic magnetic field is not strong enough to
accelerate them, so their typical velocity is expected to be of the
order of $10^{-3}c$. Therefore the astrophysical bounds on their flux
obtained from the condition of survival of the galactic magnetic field
(the Parker bound and its refinements) are {\em inapplicable} for
them. On the contrary, for the superheavy black poles their dynamics
is dominated by the galactic magnetic field, so most of them are soon
ejected out of the galaxy similar to the case of the ordinary GUT
monopoles. Thus, the low-lying black poles with the mass in the
interval $(6-3600)M_{Pl}$ and, consequently, the magnetic charge in
the range $(1-600)g_1$ seem to be more interesting experimentally.
Although the Parker-type bounds do not hold for these poles, their
flux can nevertheless be constrained at a similar level, $F \alt 10^{-
15} cm^{-2}sr^{-1}s^{-1}$, from the requirement that the galactic
density of the poles should not be unacceptably high. The best
available limits from the ionization detectors on the ultraheavy
monopole flux with velocities of the order of
$10^{-3}c$  have the same order of magnitude. For instance,
\begin{equation}
F < 5.6\times 10^{-15} cm^{-2}sr^{-1}s^{-1} \cite{macro94},
F < 2.7\times 10^{-15} cm^{-2}sr^{-1}s^{-1} \cite{imb94}
\end{equation}
Furthermore, if one takes into account the possibility of monopole
catalysis of nucleon decay (Rubakov-Callan effect)
\cite{rc}) then a somewhat stronger limit can be obtained
\cite{b} (it is, however, sensitive to the assumed catalysis
cross-section):
\begin{equation}
F < 5\times 10^{-16} cm^{-2}sr^{-1}s^{-1}.
\end{equation}
To summarize, we have examined the issue of magnetic charge 
quantization in the presence
of black holes. It was pointed out that quantization of magnetic charge
can lead to the mass quantization for magnetically charged black holes.
We have also discussed some implications for the experimental searches of
magnetically charged black holes.

This work was supported in part by a grant from National Science
Foundation, Division of International Program (U.S.-Australia
Cooperative Research). One of the authors (K.C.W) would also like to
acknowledge the hospitality at the Institut des Hautes Etudes Scientifiques
during the course of this work.

\end{document}